\documentclass[%
 reprint,
groupedaddress,
 amsmath,amssymb,
 superscriptaddress,
 aps,
 pra,
 longbibliography
]{revtex4-2}

\usepackage{graphicx}
\usepackage{dcolumn}
\usepackage{bm}
\usepackage{braket}
\usepackage{physics}
\usepackage{ulem}
\usepackage{pgfplots}
\pgfplotsset{compat=1.18}
\usetikzlibrary{calc}

\usepackage[unicode=true]{hyperref}
\usepackage{xcolor}
\usepackage[utf8]{inputenc}

\hypersetup{
  colorlinks = true, 
  linkcolor = blue, 
  filecolor = blue, 
  urlcolor = blue, 
  citecolor = blue 
}
\begin{document}

\title{Artificial-gauge-field–driven reentrant charge-density-wave phase in a three-leg
Bose-Hubbard ladder}

\author{Takayuki Yokoyama}
\email[]{yokotaka308@hiroshima-u.ac.jp}
\affiliation{Quantum Matter Program, Graduate School of Advanced Science and Engineering, Hiroshima University,
Higashihiroshima, Hiroshima 739-8530, Japan}

\author{Yasuhiro Tada}
\affiliation{Quantum Matter Program, Graduate School of Advanced Science and Engineering, Hiroshima University,
Higashihiroshima, Hiroshima 739-8530, Japan}
\affiliation{Institute for Solid State Physics, University of Tokyo, Kashiwa 277-8581, Japan}

\begin{abstract}
We investigate hard-core bosons at half filling on a three-leg ladder under the uniform artificial gauge field. By analyzing current patterns and correlation functions, we uncover a rich quantum phase diagram containing multiple superfluid and insulating phases. In bosonic ladder systems, increasing the gauge flux typically destabilizes the Meissner phase and leads to vortex phases characterized by circulating currents. In the present system, however, we find that charge-density-wave (CDW) phases emerge precisely in such a flux regime despite the presence of only an on-site interaction, where vortex states are naturally expected and are indeed realized in nearby parameter regions. While part of this behavior can be qualitatively understood from a strong-coupling perspective, we also identify an isolated CDW region that cannot be connected to such limits. Furthermore, upon increasing the artificial gauge flux, we observe a reentrant sequence of quantum phase transitions, CDW $\to$ vortex-superfluid $\to$ CDW, revealing a strong competition between the vortex phase and the density-wave order.

\end{abstract}

\maketitle

\section{\label{sec:introduction}Introduction}

Recent advances in quantum simulation platforms such as ultracold atomic systems have enabled the experimental realization of strongly correlated bosonic systems under external gauge fields \cite{science.aal3837, Schafer2020}. In particular, laser-assisted tunneling techniques in optical lattices have been developed to generate artificial gauge fields \cite{PRL.111.185301, PRL.111.185302}. These platforms have opened a door to controlled studies of gauge fields responses in low-dimensional systems.

In particular, ladder systems provide an intermediate setting between one- and two-dimensional lattices. Although their reduced geometry simplifies the problem compared with fully two-dimensional systems, they still retain essential aspects of correlated quantum many-body physics and remain highly accessible to both analytical and numerical approaches \cite{Giamarchi_book, Dagotto_1996}. In recent years, theoretical studies have revealed rich gauge-field-driven physics in bosonic-ladder systems, identifying a variety of phases, including the Meissner and vortex phases, and providing a well-developed picture of the interplay between artificial gauge fields and strong correlations \cite{PRB.64.144515, PRB.91.140406, PRB.92.060506, PRL.111.150601, PRL.115.190402, PRA.94.063628, PRA.95.063601, PRA.92.053623, Tokuno_2014, PRA.93.053629, PRB.91.054520, PRB.96.014518, PRX.7.021033, PRB.99.245101, PRR.5.013126, Kolley_2015, PRR.5.L042008, ZHANG2024130022, PRL.132.153401, PRA.113.L031301}. Experimentally, these phases have also been realized in ultracold-atom setups, where laser-assisted tunneling in optical lattices enables the observation of Meissner and vortex state in bosonic ladders \cite{Atala2014, science.aaa8515, Li2023, Impertro2025}.

The three-leg ladder moves closer to two-dimensional physics while hosting richer flux responses and potentially novel quantum phases, yet its phase diagram remains less explored. Previous numerical studies of three-leg bosonic ladders have focused mainly on the commensurate filling of one-third, where a Mott-insulating phase is allowed \cite{Kolley_2015}. At this commensurate density, a variety of phases appear such as Meissner-superfluid, Meissner-Mott states, vortex liquids (lattices), and staggered-current state. By contrast, numerical studies at half-filling remain unexplored. In this regime, a three-leg ladder with an artificial gauge field exhibits strong kinetic frustration, especially away from simple commensurate fillings. This frustration generically enhances competition among multiple ordering tendencies, giving rise to a rich and nontrivial phase structure that cannot be anticipated from the two-leg ladders.
Especially, since increasing the flux destabilizes the Meissner phase and usually promotes vortex states with circulating currents in strongly-correlated bosonic ladders, it is natural to expect that large gauge flux favor vortex-dominated phases. However, it remains unclear whether alternative ordering tendencies can compete with and even replace vortex states in such regimes.

In this work, we investigate the phase diagram of hard-core bosons at half filling on a three-leg ladder under the artificial gauge field, employing the density-matrix renormalization group (DMRG) method \cite{PRL.69.2863, SCHOLLWOCK201196, McCulloch_2007}. The obtained phase diagram differs markedly from those of the two-leg ladder and previously studied three-leg system at one-third filling. We identify a complex phase diagram characterized by multiple competing quantum phases, including current-carrying superfluid states and density-modulated phases, whose stability is strongly influenced by the interplay of the gauge field and the inter-leg coupling. Especially, we find a charge-density-wave phases for wide ranges of the gauge field, which are not usually expected in a system without an inter-site interaction.
They emerge in a parameter regime where vortex states are naturally expected and are separated by an intermediate vortex phase, indicating a direct competition between vortex formation and density ordering. These findings suggest that the three-leg ladder at half filling is a good platform for exploring quantum phases beyond the conventional mechanism induced by artificial gauge fields.

\section{\label{sec:model}Model and Method}

\subsection{Bose-Hubbard ladder with artificial gauge field}
\begin{figure}[t]
 \centering
   \includegraphics[width=1.0\linewidth]{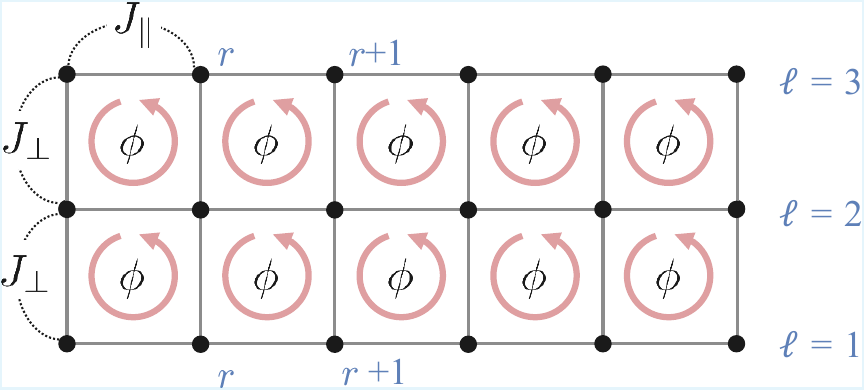}
 \caption{Schematic illustration of a three-leg ladder under an artificial gauge field.
    The sites are labeled by the rung index $r=1,2,\cdots,L$ (horizontal direction) and the leg index $\ell =1,2,3$ (vertical direction). $J_{\parallel}$ and $J_{\perp}$ denote the tunneling amplitudes along the legs and rungs, respectively. Gauge flux are inserted through each plaquette. The open boundary condition in the ladder direction has been imposed.}
 \label{fig:model}
\end{figure}
We study the Bose-Hubbard model in the presence of an artificial gauge field on the three-leg ladder as shown in Fig.~\ref{fig:model}. We focus on the half filling case, in which the numbers of boson and holes are equal corresponding to exact particle–hole symmetry ($\mathbb{Z}_2$ symmetry).
The Hamiltonian is 
\begin{align}
\mathcal{H} = &-J_{\parallel} \sum_{\ell = 1}^3 \sum_{r} \left( e^{i A_{r,\ell}} b^{\dagger}_{r+1,\ell} b^{}_{r, \ell} + \text{h.c} \right) \nonumber \\
&- J_{\perp}  \sum_{\ell = 1}^2 \left( b^{\dagger}_{r, \ell+1} b^{}_{r, \ell} 
+ \text{h.c} \right) \nonumber \\
&+ U \sum_{\ell = 1}^3 \sum_{r} n_{r, \ell} \left( n_{r, \ell} - 1 \right),
\label{eq:Hamiltonian}
\end{align}
where $b^{\dagger}_{r, \ell} (b_{r, \ell})$ is the creation (annihilation) operator for a boson on site $(r, \ell)$ and $n_{r, \ell}$ denotes the corresponding particle number. The index $r=1,2,\cdots,L$ labels the rung position and $\ell=1,2,3$ denotes the leg index. $L$ is the system length in the leg direction. $J_{\parallel}$ and $J_{\perp}$ are the tunneling amplitudes along the legs and between the rungs, respectively. We set $J_{\parallel} = 1$ as an energy unit. The on-site repulsive interaction strength $U$ can be tuned via a Feshbach resonance. In this work, we consider the hard-core boson limit $(U/J^{\parallel} \to \infty)$ where double occupancy of bosons on a site is forbidden, analogous to the Pauli exclusion principle for fermions. The vector potential $A_{r, \ell}$ is given by
\begin{equation}
A_{r, \ell} = \left\{
\begin{array}{lll}
\phi & (\ell = 1) \\
0 & (\ell = 2) \\
-\phi & (\ell = 3).
\end{array}
\right.
\label{eq:vector_potential}
\end{equation}
This configuration corresponds to a unifrom artificial gauge flux $\phi$ threading each plaquette, as realized in ultracold atomic experiments using the laser-assisted tunneling techniques \cite{PRL.111.185301, PRL.111.185302}. At half filling, in contrast to the one-third filling case, a trivial non-degenerate Mott-insulating phase is not realized. This can be understood from the Oshikawa–Yamanaka–Affleck condition \cite{PRL.78.1984}, which constrains the occurrence of gapped insulating phases in one-dimensional and quasi-one-dimensional systems with fractional filling. A detailed discussion is provided in Appendix~\ref{app:OYA}.

There are several spatial symmetries in our model (see also Appendix~\ref{app:pg}). Under the periodic boundary condition or if boundaries are neglected, the model has the translation symmetry and the $C_2$ rotation symmetry. In addition, there is a mirror symmetry combined with the particle-hole conjugation $b_{r,\ell}\to b_{r,\ell}^{\dagger}$ under the gauge field Eq.~\eqref{eq:vector_potential} in the hard-core limit $U\to\infty$~\cite{Tada_2026}. The ground state obtained by the DMRG calculations can explicitly break some of the spatial symmetries in the bulk region, from which we can identify distinct quantum phases.

\begin{figure}[t]
 \centering
   \includegraphics[width=1.0\linewidth]{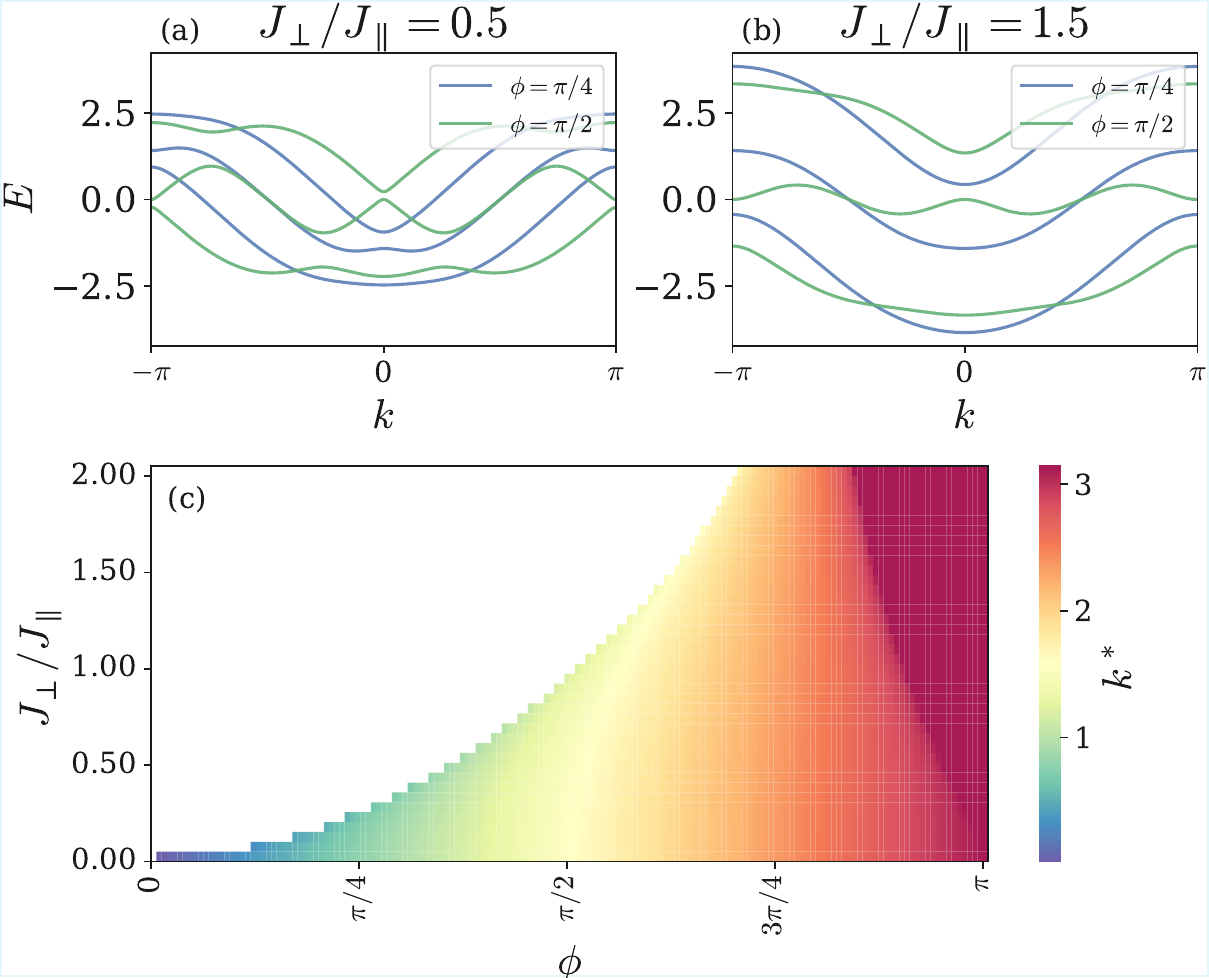}
   \caption{Non-interacting band structures and the incommensurate minimum in the three-leg ladder under an artificial gauge field. (a,b) Single-particle band dispersions for the three-leg ladder at $J_\perp/J_\parallel=0.5$ (a) and $1.5$ (b), shown for two representative gauge field flux $\phi=\pi/4$ (blue) and $\phi=\pi/2$ (green). (c) Color map of the momentum $k^\ast$ at which the lowest band attains a local minimum, excluding the minimum at $k=0$. plotted as a function of the gauge flux $\phi$ and the interleg coupling $J_\perp/J_\parallel$. White regions indicate parameter regimes where the minimum remains at $k=0$.}
 \label{fig:nonint}
\end{figure}

\subsection{Band structure in non-interacting limit}
\label{sec:noninteracting}
Before analyzing the interacting case, we first examine the non-interacting limit, which provides an important guideline for understanding the quantum phases realized in bosonic ladders with artificial gauge fields. The single-particle dispersion is closely tied to the current configurations that are stabilized once interactions are introduced. In particular, the momentum at which the lowest-energy eigenvalue is achieved often indicates the nature of the ground-state current pattern: when the minimum is at $k=0$, a translationally invariant Meissner state is expected, whereas a minimum at a finite momentum $k>0$ signals the tendency toward translational-symmetry breaking and suggests emergence of a vortex lattice in the interacting system \cite{PRA.93.053629}. A more detailed analysis of the non-interacting band structure and its relation to the current patterns has been presented in the previous work ~\cite{Sun_2020}. Here, we restrict ourselves to a brief overview of the essential features relevant to the present study.

To analyze the momentum dependence of the spectrum, we perform a Fourier transformation along the ladder direction, $b_{r,\ell} = \frac{1}{\sqrt{L}} \sum_k e^{ikr}\, b_{k,\ell}$, under the periodic boundary condition. In the non-interacting limit ($U=0$), the Hamiltonian can be written as $\mathcal{H}_0 = \sum_k \mathbf{b}_k^\dagger\, H(k)\, \mathbf{b}_k$ for  $\mathbf{b}_k^\dagger = \bigl( b_{k,1}^\dagger,\, b_{k,2}^\dagger,\, b_{k,3}^\dagger \bigr)$, where $H(k)$ is a $3\times 3$ matrix acting on the leg degrees of freedom. Since the Peierls phase is uniform along the ladder direction in our gauge condition, the flux $\phi$ enters as opposite momentum shifts on the two outer legs, leading to
\begin{equation}
  H(k) =
  \begin{pmatrix}
    -2J_{\parallel}\cos(k+\phi) & -J_{\perp} & 0 \\
    -J_{\perp} & -2J_{\parallel}\cos k & -J_{\perp} \\
    0 & -J_{\perp} & -2J_{\parallel}\cos(k-\phi)
  \end{pmatrix}.
  \label{eq:Hk_3leg}
\end{equation}
This structure highlights the essential role of the three-leg geometry: multiple dispersive channels are coupled by the inter-leg hopping $J_{\perp}$, while the artificial gauge field shifts the dispersions on the outer legs in opposite directions.

By diagonalizing $H(k)$, we obtain the single-particle bands shown in Fig.~\ref{fig:nonint}(a)(b). For small $\phi$, the lowest band exhibits a single minimum at $k=0$, consistent with a Meissner-type current configuration. As $\phi$ is increased, additional local minima appear at finite momenta $k>0$. This behavior suggests that increasing the gauge flux can drive a transition from a Meissner phase to a vortex phase. Furthermore, increasing the inter-leg coupling $J_{\perp}$ enlarges the parameter region where the minimum remains at $k=0$, while simultaneously generating metastable minima near $k\simeq \pi$ (Fig.\ref{fig:nonint}(c)). This tendency implies that stronger inter-leg coupling destabilizes vortex configurations and favors states with a larger effective vortex density. Overall, the non-interacting band structure provides a useful qualitative guideline for understanding the interacting phase diagram discussed in the following sections.

\subsection{How to characterize various phase}
\label{sec::obs}
The interacting hard-core bosons are investigated with use of DMRG for the system under the open boundary condition. We examine some phases such as Meissner phase and vortex phase based on local currents, defining $j \equiv \frac{\partial{\mathcal{H}}}{\partial{A}}$.
The local currents on legs and rungs are respectively given by
\begin{align}
j^{\parallel}_{r, \ell} &=
-i J_{\parallel} \left( e^{i A_{r, \ell}} b^{\dagger}_{r+1, \ell} b_{r, \ell} - e^{-i A_{r, \ell}} b_{r+1, \ell} b^{\dagger}_{r, \ell} \right), \\
j^{\perp}_{r, \ell} 
&= -i J_{\perp} \left( b^{\dagger}_{r, \ell+1} b_{r, \ell} -  b_{r+1,\ell} b^{\dagger}_{r, \ell} \right). 
\end{align}
We define the chiral current as the order parameter of the Meissner phase (Fig.\ref{fig:schematic}(a)):
\begin{align}
\mathcal{J}_{\text{c}} = \frac{1}{N_\mathrm{leg}} \sum_{r} \ev{j^{\parallel}_{r, 1} - j^{\parallel}_{r, 3}}. 
\label{eq:Jc}
\end{align}
So as to suppress contributions from edges of the system under the open boundary condition, we evaluate $\mathcal{J}_{\text{c}}$ using the middle region $L/4<r \le 3L/4$, for which the normalization factor is $N_{\text{leg}}=3(L-1)/2$. In few-leg ladder systems, the Meissner phase can be characterized by the orbital-diamagnetic magnetization, which is proportional to the chiral current $\mathcal{J}_{\mathrm{c}}$.

We also evaluate the averaged rung current amplitude,
\begin{align}
|\mathcal{J}_{\mathrm{rung}}|
= \frac{1}{N_{\mathrm{rung}}}
\sum_{\ell = 1}^{2}\sum_{r}
\left|\langle j^{\perp}_{r,\ell}\rangle\right|,
\label{eq:Jr}
\end{align}
which is known to serve as an effective order parameter for vortex phases in ladder systems \cite{PRB.91.140406, Kolley_2015, PRR.5.013126}, since finite rung currents emerge when vortices penetrate the ladder (Fig.\ref{fig:schematic}(b)-(d)). Similarly to the Meissner current $\mathcal{J}_{\mathrm c}$, we take only the $N_{\mathrm{rung}}=L$ sites in $L/4 < r \le 3L/4$ into account for the calculation of $\mathcal{J}_{\mathrm{rung}}$. To further diagnose vortices, we introduce the Fourier-transform of the rung current along the leg direction, $j^{\perp}_{\ell}(k) \equiv \frac{1}{\sqrt{L}} \sum_{r=1}^{L} e^{-ikr}\,\langle j^{\perp}_{r,\ell}\rangle$ and the corresponding rung-current structure factor
\begin{align}
S_{\mathrm{rung}}(k) \equiv \frac{1}{2}\sum_{\ell=1}^{2}\left|j^{\perp}_{\ell}(k)\right|^{2}.
\end{align}
A peak in $S_{\mathrm{rung}}(k)$ at a finite wave number $k=k_{\mathrm{v}}$ signals a vortex pattern with the period $2\pi/k_{\mathrm{v}}$ along the ladder. 

At one-third filling, a phase with staggered currents emerges in the presence under the artificial gauge field (Fig.\ref{fig:schematic}(d))~\cite{Kolley_2015}. To characterize this state also in the present half filling case, we define the staggered current along the legs as
\begin{align}
    \mathcal{J}_{\mathrm{sc}}
    = \frac{1}{N_{\text{leg}}}\sum_{r,\ell}
    (-1)^{\ell + r}\,\langle j^{\parallel}_{r,\ell} \rangle.
    \label{Jsc}
\end{align}

The essential distinction between the vortex state and the staggered-current state lies in their symmetry properties (see also Appendix~\ref{app:pg}). While the vortex phases break translational symmetry along the leg direction due to the formation of circulating current patterns, they preserve the bond-centered $C_2$ rotation symmetry. In contrast, the staggered-current phase breaks the bond-centered $C_2$ rotation symmetry. Therefore, although both phases exhibit non-vanishing currents induced by the gauge flux, the staggered-current phase should be regarded as a distinct current-ordered state, rather than a variant of the vortex phase.

We characterize the superfluid and vortex phases by introducing two-point correlation functions defined as a function of the distance $r$ between the sites $(r_1,\ell)$ and $(r_2,\ell)$ along the ladder with $r_1 =L/2$ and $r_2=L/2+r$. Specifically, we define the superfluid correlation function and the
current-current correlation function as
\begin{align}
C^{\mathrm{SF}}_{r} 
&= \frac{1}{3} \sum_{\ell = 1}^3
\left| \left\langle  b^{\dagger}_{r_1,\ell}\, b_{r_2,\ell} \right\rangle \right|,
\label{eq:SF_corr} \\
C^{\mathrm{V}}_{r}
&=
 \frac{1}{2} \sum_{\ell = 1}^2
\left|  \left\langle  j^{\perp}_{r_1,\ell}\, j^{\perp}_{r_2,\ell} \right\rangle \right|,
\label{eq:V_corr}
\end{align}
where the correlations are averaged over the leg index $\ell$. The superfluid phase is a quasi long range order with gapless excitations in a quasi one-dimensional system. As a result, the superfluid correlation function $C^{\mathrm{SF}}_{r}$ exhibits a power-law decay with distance and $C_r^{\mathrm{V}}$ is exponentially suppressed in the superfluid phase. On the other hand, in the vortex phase of a two-leg ladder system, the current-current correlation function $C^{\mathrm{V}}_{r}$ also shows a power-law decay. Therefore, $C^{\mathrm{V}}_{r}$ serves as a useful diagnostic for identifying the vortex state in ladder systems.

The staggered CDW order parameter is
\begin{align}
\mathcal{O}_{\mathrm{CDW}} = \frac{1}{N_\mathrm{site}}\sum_{\ell = 1}^3 \sum_{r} (-1)^{r+ \ell} \ev{n_{r, \ell}},
\label{eq:CDW}
\end{align}
where $N_\mathrm{site}=3L/2$ which is the number of sites in the middle region $L/4<r \le 3L/4$. The expectation value of this operator becomes nonzero in the CDW phase under the open boundary condition (Fig.\ref{fig:schematic}(f)).

\subsection{Details of numerical annalysis}
In this work, we perform DMRG calculations using \textit{ITensor} \cite{ITensor}, which is based on the matrix product state (MPS) formalism and is well suited for analyzing quasi-one-dimensional systems such as ladders. We compute the ground state of the hard-core bosonic ladder for a system length of $L=80$ under the open boundary condition. We use the maximum bond dimension $\chi=1000$. The resulting truncation error $\varepsilon$ typically satisfies $10^{-8}\lesssim \varepsilon \lesssim 10^{-5}$, which is sufficiently small for determining the phase diagram.

\section{Numerical results}
\begin{figure}[t]
 \centering
   \includegraphics[width=1.0\linewidth]{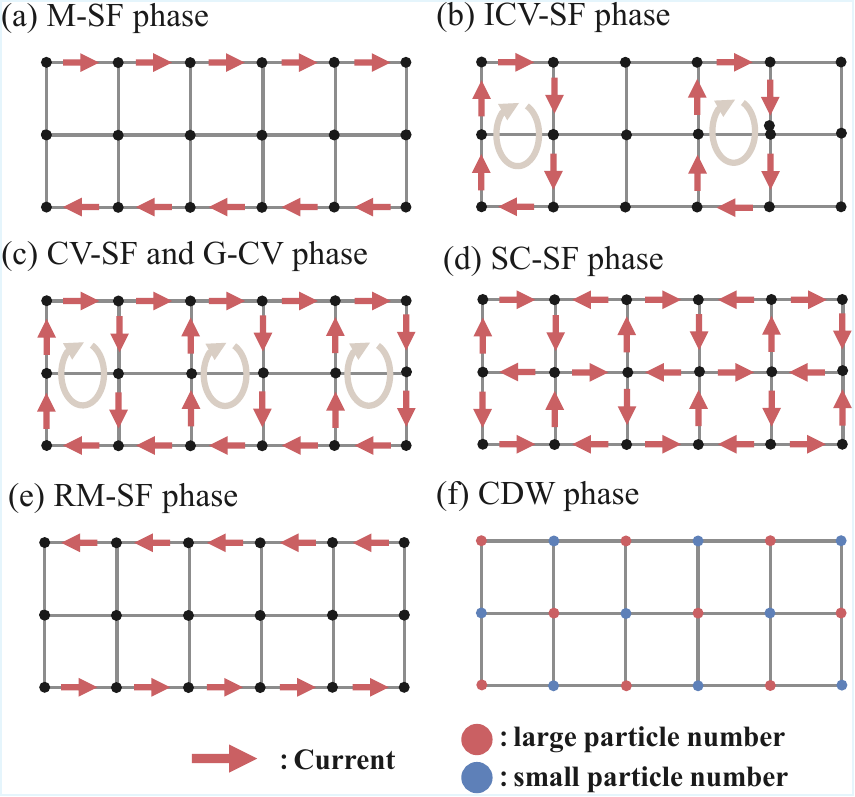}
 \caption{Schematic illustrations of current and density configurations in the different phases of the three-leg ladder. Red arrows indicate the direction of particle currents. Panels show (a) the Meissner superfluid (M-SF) phase, (b) the incommensurate vortex-superfluid (ICV-SF) phase, (c) the commensurate vortex-superfluid (CV-SF) and gapped commensurate vortex (G-CV) phases, (d) the staggered-current superfluid (SC-SF) phase, (e) the reverse-Meissner-superfluid (RM-SF) phase, and (f) the charge-density-wave (CDW) phase. In the CDW phase, red and blue circles represent sites with large and small particle densities, respectively.
}
 \label{fig:schematic}
\end{figure}

\begin{figure}[t]
  \centering
  \includegraphics[width=1.0\linewidth]{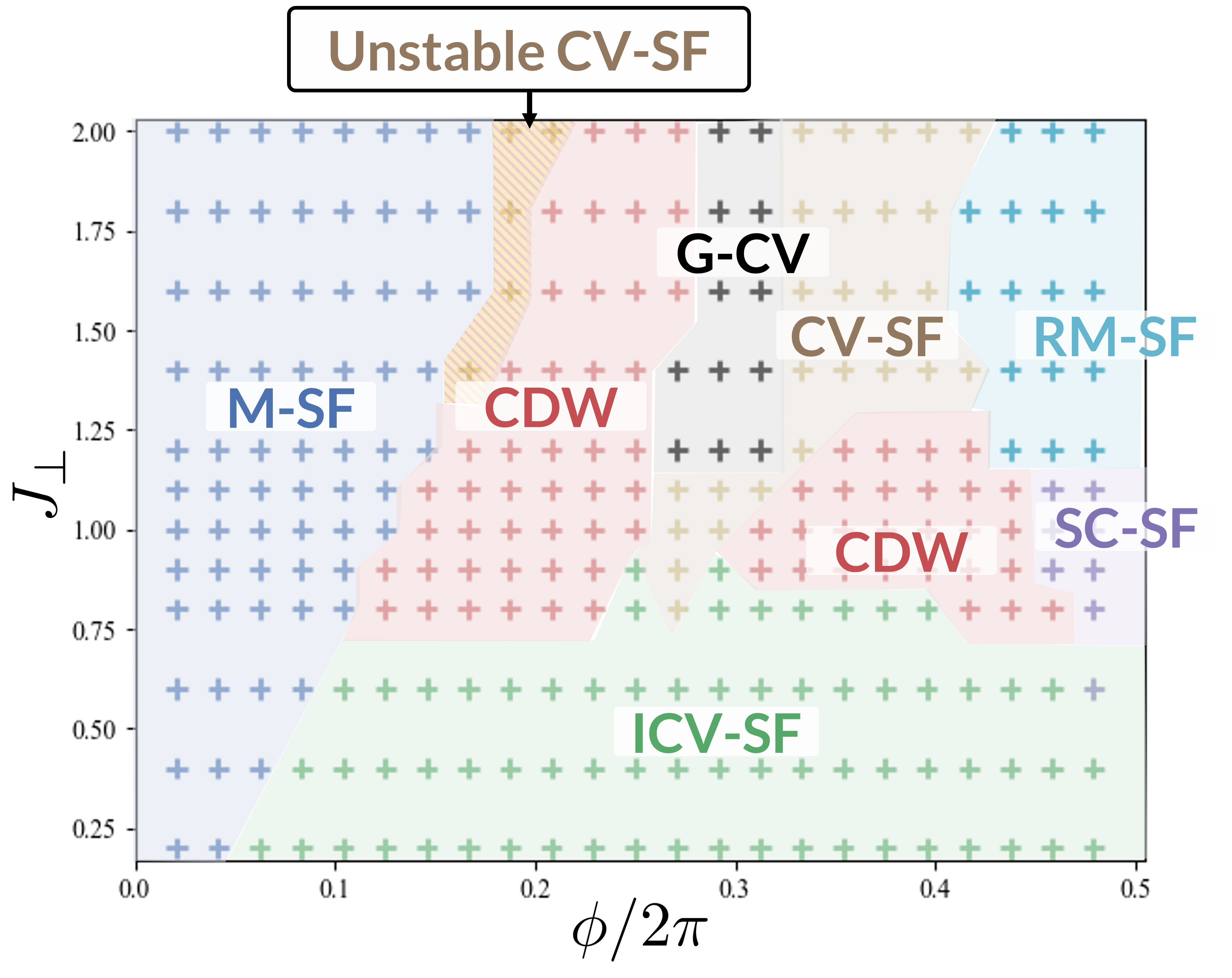}
  \caption{Phase diagram for $L=80$ as a function of the flux $\phi$ and the interleg coupling $J_\perp$ (in the unit of $J_{\parallel} = 1$). Distinct phases are identified, including the Meissner superfluid (M-SF) phase, incommensurate vortex superfluid (ICV-SF) phase, commensurate vortex superfluid (CV-SF) phase, unstable CV-SF phase, gapped commensurate vortex phase (G-CV) phase, staggered-current superfluid (SC-SF) phase, reverse-Meissner-superfluid (RM-SF) phase and charge-density-wave (CDW) phase.}
  \label{fig:phase_diagram}
\end{figure}
\subsection{Overview of the phase diagram}
\begin{table*}[t]
\centering
\renewcommand{\arraystretch}{1.25}
\setlength{\tabcolsep}{10pt}
\caption{Summary of the characteristic properties of the quantum phases. The terms "power-law" and "exponential" denote the spatial decay of the superfluid correlation $C_r^{\mathrm{SF}}$, and a dash "--" in $\mathcal{J}_c$ indicates that it is not relevant for characterizing the phase. In the SC-SF phase, "$\text{No vortex}^*$" indicates that circulating currents are present, but they are not vortex state such as a type-II superconductor with a magnetic field.}
\label{tab:phase_properties}
\begin{tabular}{lcccccc}
\hline\hline

& $C^{\text{SF}}_r$
& $\mathcal{J}_{\text{c}}$
& $|\mathcal{J}_{\text{rung}}|$
& $|\mathcal{J}_{\text{SC}}|$ 
& Vortex structure
& $\mathcal{O}_{\text{CDW}}$ \\
\hline
M-SF
& Power-law 
& $>0$
& 0
& 0 
& No vortex
& 0
\\

CDW  
& Exponential 
& -
& $0$ 
& 0
& No vortex
& $>0$
\\

ICV-SF
& Power-law 
& $>0$
& $>0$ 
& 0
& Incommensurate
& 0
\\

CV-SF
& Power-law 
& -
& $>0$ 
& 0
& Commensurate
& $0$
\\

G-CV
& Exponential 
& -
& $>0$ 
& 0
& Commensurate
& $0$
\\

SC-SF
& Power-law 
& -
& $>0$ 
& $>0$
& $\text{No vortex}^*$
& 0
\\

RM-SF
& Power-law 
& $<0$
& 0 
& 0
& No vortex
& 0
\\

\hline\hline
\end{tabular}
\end{table*}
Figure~\ref{fig:phase_diagram} is the phase diagram of the hard-core bosons at half filling on the three-leg ladder under the artificial gauge field. Several distinct phases are identified: the Meissner-superfluid phase (M-SF), incommensurate-vortex-superfluid phases (ICV-SF), commensurate-vortex-superfluid phase (CV-SF), gapped-commensurate-vortex phase (G-CV), charge-density-wave phase (CDW), staggered-current-superfluid phase (SC-SF), and reverse-Meissner-superfluid phase (RM-SF). A schematic illustration of these phases is shown in Fig.~\ref{fig:schematic}, while their characteristic properties are summarized in Table~\ref{tab:phase_properties}. The M-SF phase is characterized by the power-law–decaying superfluid correlations (Eq.~\eqref{eq:SF_corr}) and by the edge chiral currents (Eq.\eqref{eq:Jc}) flowing in the diamagnetic direction. This phase is also realized in two-leg bosonic ladder systems on square and triangular lattice geometries \cite{PRL.111.150601, PRR.5.013126}. The ICV-SF and CV-SF phases both exhibit quasi-long-range superfluid correlations accompanied by a periodic vortex structure, reminiscent of a vortex lattice. The essential distinction between the two phases lies in the periodicity of the vortex pattern: while the ICV-SF phase is characterized by an incommensurate vortex modulation, the CV-SF phase displays a commensurate vortex lattice locked to the underlying lattice. The G-CV phase is a gapped insulating phase characterized by the absence of superfluid correlations. Nevertheless, the current pattern exhibits a periodic vortex structure, indicating the persistence of vortex ordering without superfluid coherence. We also find a finite-size-unstable CV-SF region, whose stability in the thermodynamic limit ($L \to \infty$) is subtle within the present calculations. (See the App .~\ref{app:finite_size}.) The CDW phase corresponds to a solid-like density pattern, analogous to a checkerboard structure, and is characterized by the order parameter $\mathcal{O}_{\mathrm{CDW}}$ (Eq.~\eqref{eq:CDW}). Remarkably, the emergence of this phase is nontrivial: in conventional systems, the CDW order is typically absent when there is no inter-site interaction (such as a nearest-neighbor interaction). The SC-SF phase represents superfluid-correlation and is characterized by circulating currents that alternate their directions on neighboring plaquettes, and it persists even at one-third filling. \cite{Kolley_2015}. The RM-SF phase is a superfluid phase in which the direction of the chiral current is reversed compared to that of the M-SF phase.
\begin{figure}[t]
  \centering
  \includegraphics[width=1.0\linewidth]{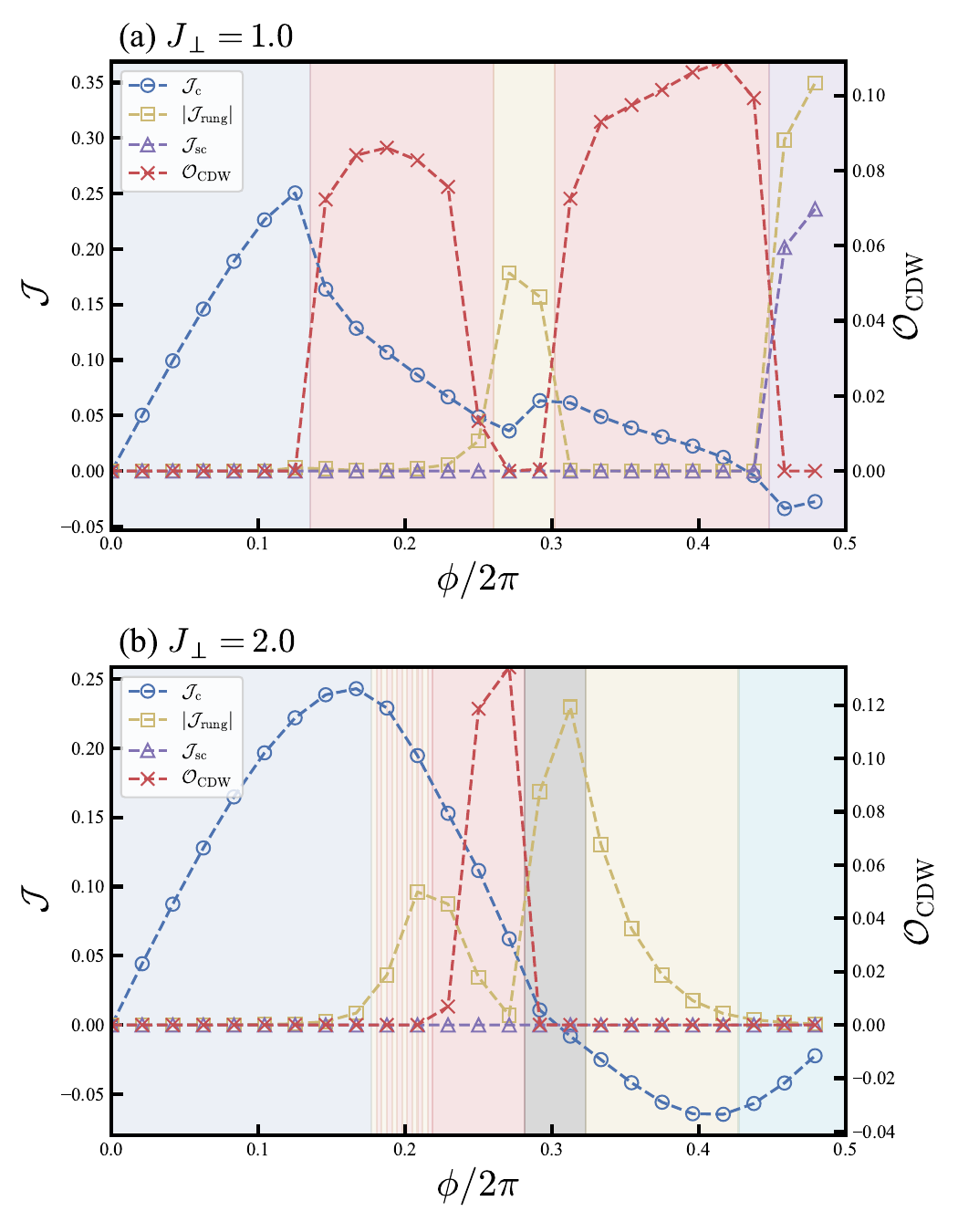}
  \caption{Gauge flux dependence of the chiral current $\mathcal{J}_c$, the rung current amplitude $|\mathcal{J}_{\mathrm{rung}}|$, the staggered current $\mathcal{J}_{\text{sc}}$, and the CDW order parameter $\mathcal{O}_{\mathrm{CDW}}$ for the three-leg hard-core bosonic ladder. Results are shown for (a) $J_\perp =1.0$, (b) $J_\perp =2.0$. Shaded regions indicate the various quantum phases. The blue region is the M-SF phase, where $\mathcal{J}_c$ increases monotonically with $\phi$. The red regions is CDW phase. The yellow region is the CV-SF phase, and the black region is the G-CV phase. Both phases exhibit a finite rung current $\mathcal{J}_{\mathrm{rung}}$. Their distinction is discussed in Sec.~\ref{sec:ICV-SF_CV-SF} and Sec.\ref{sec:G-CV}. The light blue region is the RM-SF phase, where $\mathcal{J}_c$ has the opposite sign to that in the M-SF phase. The striped region indicates the finite-size-unstable CV-SF region.
}
  \label{fig:phase_diagram_Jrfix}
\end{figure}
In conventional superfluids, magnetic flux (artificial gauge flux) drives a quantum phase transition from the M-SF phase to the V-SF phase.
In our model, for $J_{\perp} \simeq 1.0$, the vortex phases are separated from the M-SF phase by an intermediate CDW phase, and there is another CDW phase between the vortex phase and the SC phase. This is a rather unconventional behavior, absent not only in the three-leg ladder at one-third filling \cite{Kolley_2015}, but also in any other system.

We note that the phase diagram in Fig.~\ref{fig:phase_diagram} is obtained from finite-size DMRG calculations for $L=80$. Therefore, finite-size effects should be carefully considered, especially for narrow phase regions such as CV-SF phases. To reveal stability of this phase in the thermodynamics limit, we performed DMRG calculations for several system sizes along representative cuts of the phase diagram. As discussed in Appendix~\ref{app:finite_size}, we show that the CV-SF phase for $J_\perp \simeq 1.0$ is stable and is not finite-size artifacts of the DMRG calculations.

\subsection{Unexpected CDW phases}
\label{sec:CDW}
Charge-density-wave (CDW) order is usually associated with inter-site interactions. Therefore, in the Bose-Hubbard model with only on-site interactions, one might naively expect that it is difficult to stabilize CDW phases. Nevertheless, previous studies have shown that artificial gauge fields can induce density-modulated states in the hard-core bosonic two-leg ladder without an explicit inter-site interaction \cite{PRB.91.140406, PRA.94.063628}.
In the present three-leg ladder at half filling, we also find CDW states in the phase diagram (red regions in Fig.~\ref{fig:phase_diagram_Jrfix}(a)(b)), although increasing the flux typically leads to vortex phases characterized by circulating currents in bosonic ladders. The CDW order parameter (Eq.\eqref{eq:CDW}) is nonzero and the superfluid off-diagonal correlation function $C^{\text{SF}}_r$ (Eq.~\eqref{eq:SF_corr}) decays exponentially with distance in this region (Fig.~\ref{fig:correlation}(d)).

In our model, two distinct CDW phases emerge: one connected to the strong-rung regime and another appearing as an isolated island in the intermediate-coupling region. We emphasize that they emerge in regimes where the Meissner states are destabilized by the flux, vortex states are naturally expected and indeed realized in nearby parameters. The former can be naturally interpreted as an extension of the previously discussed mechanism~\cite{PRB.91.140406, PRA.94.063628}. In particular, the CDW phase connected to the strong rung-coupling regime is consistent with the idea that the gauge flux and hopping processes can induce an effective coupling which favors density ordering even in the absence of explicit inter-site interactions. We discuss this picture based on an perturbative analysis in Sec.~\ref{sec:strong_coupling}. However, the perturbative argument alone is not sufficient to fully account for the stability of this phase. In our numerical results, the CDW order remains robust even around $J_{\perp} \simeq J_{\parallel}$, well beyond the regime where the strong-rung-coupling picture is expected to be quantitatively reliable. This indicates that the CDW phase in the present system is considerably more robust and nontrivial than what would be anticipated from a simple strong-coupling argument. In addition, there is almost no vortex-like current when $J_{\perp} \simeq J_{\parallel}$ in contrast to the naive expectation for bosonic ladders under fluxes. An even more remarkable feature is the emergence of an isolated island-like CDW region adjacent to the SC-SF phase. Because this phase is not continuously connected to the strong-rung CDW phase, its origin cannot be understood by the same perturbative mechanism. Rather, it should be viewed as a non-perturbative many-body phase stabilized by the competition between vortex formation and density ordering in the intermediate regime.

In the vicinity of $J_{\perp} \simeq 1$, the existence of the two distinct CDW phases lead to a reentrant phase transition of the form $\text{CDW} \rightarrow \text{CV-SF} \rightarrow \text{CDW}$ upon increasing the gauge flux, as shown in Fig.~\ref{fig:phase_diagram} and Fig.~\ref{fig:phase_diagram_Jrfix}(a). This behavior points to a delicate energetic competition between the CV-SF and CDW phases. Rather than acting as a simple tuning parameter that monotonically favors one order over the other, the gauge flux can stabilize distinct ordered states in different parameter regimes. This reentrant structure provides further evidence that the interplay between vortex physics and density ordering in the present system is not straightforward.

A more detailed analysis of this reentrant transition is presented in Appendix~\ref{app:finite_size}. We find abrupt changes in the CDW order parameter across the transition between the CDW and CV-SF phases. This suggests that the transition between the CDW and CV-SF phases is unlikely to be described as continuous transition, but rather reflects a direct competition between two distinct ordered states.

Finally, the stability of the CDW phases depends sensitively on the filling fraction. In the present system, the staggered CDW order with wave vector $k=\pi$ is compatible with half filling, since a period-two modulation preserves the average density. However, this commensurability alone does not uniquely select the observed CDW pattern, as other density configurations can also be consistent with half filling. This indicates that the emergence of the staggered CDW phase is not solely determined by the filling condition, but detailed structure is determined by the competition between vortex physics and density ordering

\subsection{M-SF phase and RM-SF phase}
\label{sec:M-SF_and_RM-SF}
\begin{figure*}[t]
  \centering
  \includegraphics[width=1.0\textwidth]{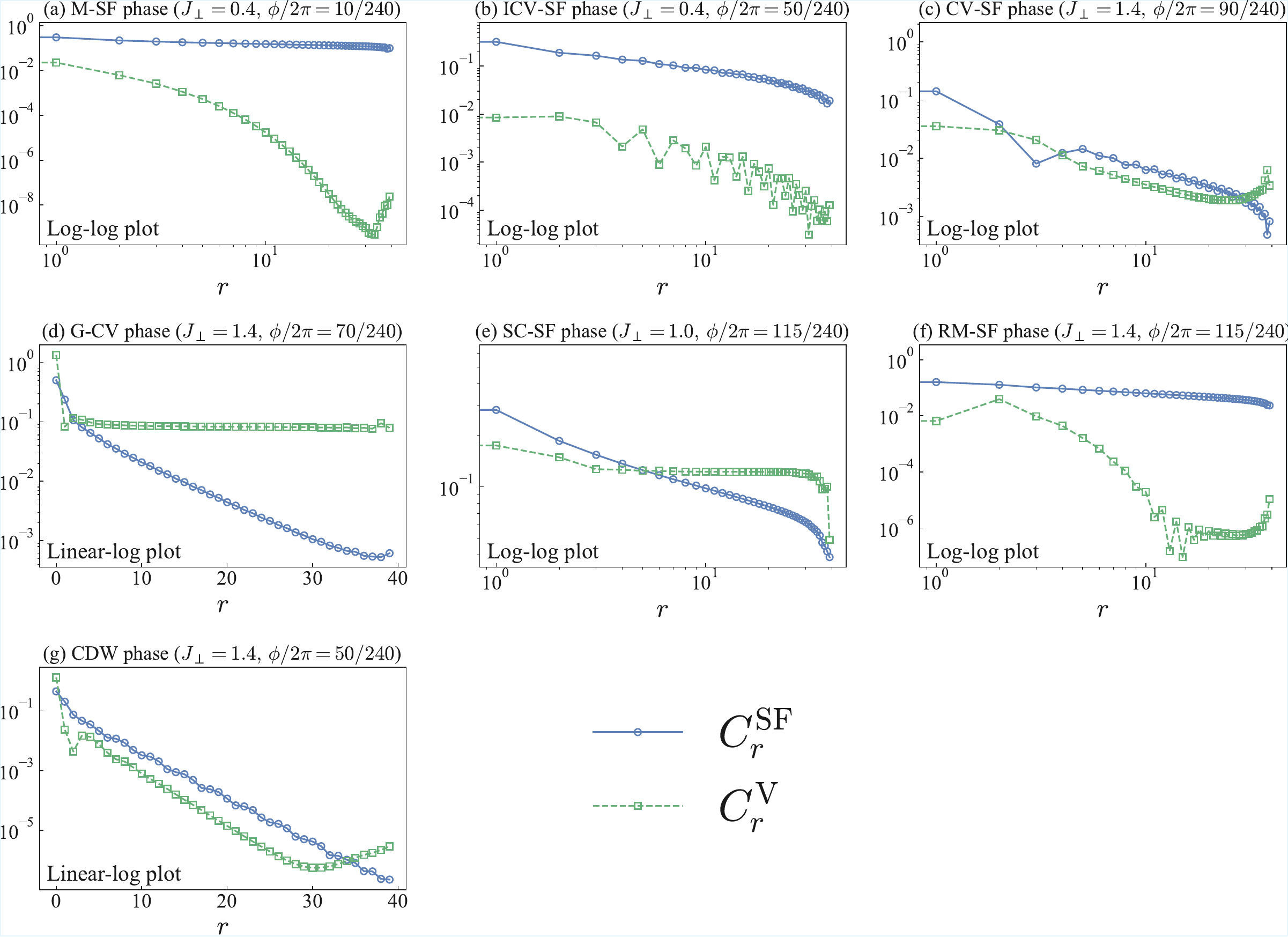}
  \caption{Superfluid correlation functions. $C^{\text{SF}}_r$ (blue circles) and the magnitude of the current-current correlation $C^{\text{V}}_r$ (green squares), plotted as functions of the distance $x$ along the leg direction between the sites $r_1=L/2$ and $r_2=L/2+r$ (Eq.\eqref{eq:SF_corr}, Eq.\eqref{eq:V_corr}). Figures (a)–(g) correspond to representative various parameter points. Log–log or linear–log scales are used depending on the panel, as annotated, to highlight the distinct decay behaviors of correlations in different phases. Note that the upturn behavior around $r \simeq L/2=40$ is a boundary effect}
  \label{fig:correlation}
\end{figure*}
The Meissner phase is characterized by a finite chiral current $\mathcal{J}_{\text{c}}$  and absence of the rung current $|\mathcal{J}_{\text{rung}}|$ (Eqs.~\eqref{eq:Jc} and \eqref{eq:Jr}). Figure~\ref{fig:phase_diagram_Jrfix}(a)(b) shows the chiral and rung currents for $J_\perp = 1.0$, and 2.0, with the chiral current plotted in blue and the rung current in yellow. For a small gauge flux $\phi$, the chiral current increases linearly, while the rung current remains zero, which is characteristic of the diamagnetic response like the Meissner effect in superfluids (superconductors). We also find that the superfluid off-diagonal correlation function $C^{\text{SF}}_r$ (Eq.\eqref{eq:SF_corr}) exhibits a quasi long-range order (Fig.~\ref{fig:correlation}(a)) and therefore identify this chiral current phase at small $\phi$ as the Meissner superfluid (M-SF) phase. This phase is a gapless Tomonaga-Luttinger liquid, which is further supported by the behavior of the entanglement entropy (see Appendix.\ref{app:enta}).

We remark that the present M-SF phase should be distinguished from a conventional Meissner phase in two-dimensional superconductors (superfluids) with magnetic fields. A Hamiltonian formulated with a given Peierls phase describes a system in which the gauge flux (magnetic flux) penetrates the bulk. Consequently, there is a diamagnetic current under the fixed gauge flux. In contrast, the true Meissner phase is characterized by the complete expulsion of an applied magnetic field whose classical dynamics is described by the Maxwell's equation. 

In contrast to the M-SF phase at small $\phi$, for the coupling $J_{\perp} \gtrapprox 1.0$ and flux $\phi$ near $\pi$, the chiral current becomes negative while the rung current remains zero (Fig.\ref{fig:phase_diagram_Jrfix}(b)). This behavior corresponds to a reverse-Meissner-current, which was previously identified at one-third filling in the three-leg hard-core bosonic ladder \cite{Kolley_2015}. The superfluid off-diagonal correlation function $C^{\text{SF}}_r$ exhibits a power-law decay, signaling quasi-long-range superfluid order (Fig.\ref{fig:correlation}(f)). We thus identify this phase as the reverse-Meissner-superfluid (RM-SF) phase.

\subsection{ICV-SF phase and CV-SF phase}
\label{sec:ICV-SF_CV-SF}
\begin{figure*}[t]
  \centering
  \includegraphics[width=1.0\textwidth]{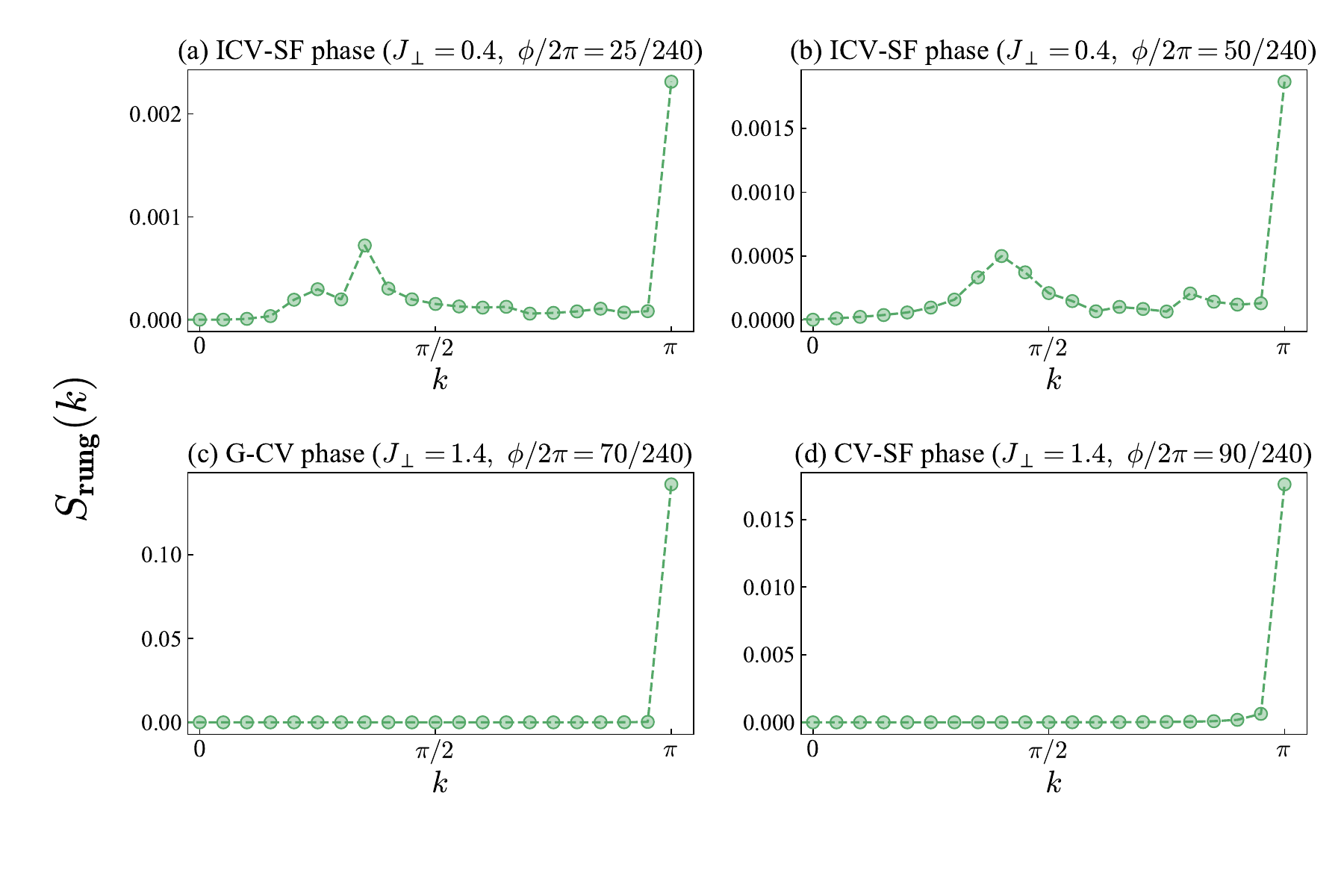}
  \caption{Momentum-resolved rung-current structure factor $S_{\mathrm{rung}}(k)$. Figure (a) and (b) show results in the incommensurate vortex superfluid (ICV-SF) phase at $J_\perp=0.4$ with $\phi/2\pi=25/240$ and $\phi/2\pi=50/240$, respectively, where only weak and broad peaks are visible, reflecting where weak and broad peak are visible away from $k=\pi$, indicating the presence of incommensurate vortex configuration. Figure (c) and (d) correspond to the gapped-commensurate vortex (G-CV) phase at $J_\perp=1.4$, $\phi/2\pi=70/240$ and the commensurate vortex superfluid (CV-SF) phase at $J_\perp=1.4$, $\phi/2\pi=90/240$, respectively. In contrast to the ICV-SF phase, the G-CV and CV-SF phases exhibit a pronounced enhancement at $k=\pi$, signaling the emergence of commensurate vortex lattice. Note the difference in the vertical axis scale.}
  \label{fig:current}
\end{figure*}

Upon increasing the gauge flux $\phi$ from the M-SF phase, vortices are induced through several intermediate phases, which are characterized by a finite magnitude of the rung current serving as a useful order parameter in bosonic ladders. We find an intermediate regime in which the rung current becomes finite as the gauge flux $\phi$ is increased, corresponding to the yellow regions in Figs.~\ref{fig:phase_diagram_Jrfix}(a)(b). In addition, the power-law-decay of $C^{\text{SF}}_r$ (Eq.\eqref{eq:SF_corr}) further supports the superfluid character of this phase (Fig.\ref{fig:correlation}(b)(c)). This phase can naively be regarded as a vortex-superfluid phase. We also find a unstable vortex region near $J_\perp \simeq 1.8$, which might vanish in the thermodynamic limit ($L \to \infty$), as discussed in Appendix~\ref{app:finite_size}.

The vortex-superfluid phases can be broadly classified into two distinct types, depending on whether the vortex configuration is commensurate or incommensurate with the lattice structure of the system. One characteristic feature of the incommensurate vortex-superfluid (ICV-SF) phase is that the periodicity of the vortex configuration continuously changes as the artificial gauge flux $\phi$ is increased. Figure~\ref{fig:current}(a)(b)(d) shows the Fourier spectrum of the rung current in the ICV-SF phase and CV-SF phase. In the CV-SF phase, we find a single peak. In contrast, the ICV-SF phase exhibits two distinct peaks, indicating the coexistence of two characteristic length scales associated with vortex correlations. While the mode at $k=\pi$ is commensurate with the lattice for $J_{\perp} \gtrsim 1.0$ (CV-SF phase), the additional mode at $k\neq\pi$ is incommensurate for $J_{\perp} < 1.0$ (ICV-SF phase). The peak position of the additional mode depends on the applied flux $\phi$.

It is important to note that the quantum phases are distinguished by the direction of vortex circulation, namely whether the vortices exhibit a diamagnetic or paramagnetic response. In a conventional type-II superfluid with a magnetic flux, both of the Meissner phase and the vortex phase show diamagnetic responses. The ICV-SF phase observed here follows this conventional type-II–like behavior. By contrast, the CV-SF phase does not necessarily fall into this category. As illustrated in Fig.~\ref{fig:phase_diagram_Jrfix}(b), the chiral current in the CV-SF phase can eventually flow in a paramagnetic direction, indicating a response that is not simply understood within the conventional type-II framework. This suggests that the CV-SF phases identified in the present work should be clearly distinguished from conventional Type-II superfluids.

\subsection{G-CV phase}
\label{sec:G-CV}
The gapped commensurate vortex (G-CV) phase emerges for $J_{\perp} \gtrsim 1.0$ in an intermediate region.
This phase is characterized by a vortex configuration that is commensurate with the underlying lattice, while exhibiting suppressed superfluid correlations. In contrast to the CV-SF phase, the superfluid off-diagonal correlation function $C^{\text{SF}}_r$ (Eq.\eqref{eq:SF_corr}) decays exponentially and  $C^{\text{V}}_r$ (Eq.\eqref{eq:V_corr}) rapidly converges to a finite value (Fig.\ref{fig:correlation}(d)), 
indicating the presence of a finite excitation gap. This gapped nature is further supported by the saturation behavior of the entanglement entropy (see Appendix~\ref{app:enta}).

Despite the absence of superfluid correlations, the rung current remains finite in this phase, signaling the persistence of vortex ordering. Consistently, the Fourier spectrum of the rung current exhibits a pronounced peak at $k=\pi$ (Fig.~\ref{fig:current}(c)), demonstrating that the vortex configuration is commensurate with the lattice. A similar gapped commensurate vortex phase has been identified at one-third filling in the previous work \cite{Kolley_2015}. This observation suggests that the emergence of the G-CV phase may be a common consequence of artificial gauge fields in three-leg bosonic ladders, rather than a feature unique to the half-filled case considered here.

\subsection{SC-SF phase}
\label{sec:SC-SF}
We identify a distinct current configuration that is different from both the Meissner state and the vortex state. Although the phases discussed so far preserve a reflection symmetry, we find that, for a sufficiently large flux $\phi$ and for comparable rung and leg couplings, this symmetry is spontaneously broken and a staggered-current-superfluid (SC-SF) phase emerges. In the SC-SF phase, loop currents circulate with alternating chirality on adjacent plaquettes, leading to a doubling of the unit cell along the leg direction. This breaks the translational symmetry and the bond-centered $C_2$-rotation symmetry. It is therefore distinct from the M-SF phase, which preserves translational symmetry, and from the vortex phase, which breaks translational symmetry while keeping the $C_2$-rotation symmetry.

For $J_{\perp} \approx 1.0$ and large artificial gauge fluxes $\phi \simeq \pi$, we find that although the magnitude of the rung current remains finite, the leg currents exhibit a staggered pattern, as shown in Fig.~\ref{fig:phase_diagram_Jrfix}(b). The SC-SF phase can be characterized by two order parameters: the staggered chiral current $\mathcal{J}_{\mathrm{sc}}$ (Eq.\eqref{Jsc}), which captures the alternating loop-current pattern, and the finite magnitude of the rung current. We also find that the superfluid correlation function $C^{\text{V}}_{\text{SF}} $ (Eq.\ref{eq:SF_corr}) exhibits a power-
law decay (Fig.\ref{fig:correlation}(e)). Although circulating currents are present, they are not a vortex lattice phase such as in a type-II superconductor with a magnetic field. Instead, this phase is more appropriately understood as a loop-current-ordered state. We note that the emergence of staggered currents near $\phi \simeq \pi$ is a feature that also appears at one-third filling in the three-leg ladder~\cite{Kolley_2015}, indicating that this phase is a generic consequence in three-leg ladder systems.

\section{Discussion in strong rung-coupling limit }
\label{sec:strong_coupling}
\begin{figure}[t]
 \centering
   \includegraphics[width=1.0\linewidth]{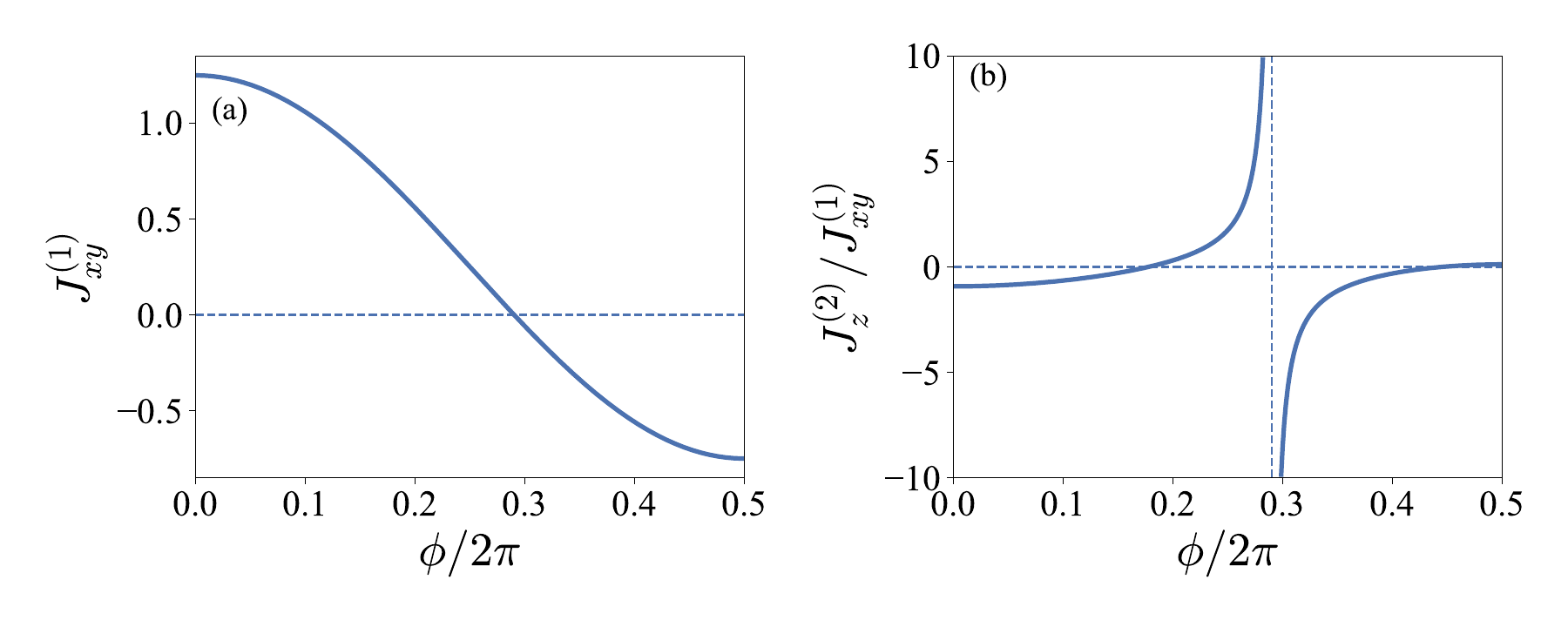}
\caption{
Effective couplings obtained from perturbation theory in the strong-rung-coupling limit.
(a) First-order XY exchange $J_{xy}^{(1)}(\phi)$ as a function of the gauge flux $\phi$. The dashed horizontal line indicates $J_{xy}^{(1)}=0$, where the sign of the effective XY coupling changes.
(b) Ratio $J_z^{(2)}(\phi)/J_{xy}^{(1)}(\phi)$. The horizontal dashed line indicates $J_z^{(2)}/J_{xy}^{(1)}=0$, while the vertical dashed line marks the point where $J_{xy}^{(1)}(\phi)=0$.}
 \label{fig:perturbation}
\end{figure}
The microscopic origin of the CDW phase is not immediately obvious from the original Hamiltonian since there is no explicit nearest-neighbor density interaction that would directly favor a density-modulated state. To gain insight into this issue, we follow the strong-coupling approach adopted in Refs.~\cite{PRB.91.140406,PRA.94.063628} and consider the limit $J_{\perp} \to \infty$, where an effective low-energy description can be derived within a perturbation theory.

In the limit $J_{\parallel}/J_{\perp} \ll 1$, each rung can be treated as an isolated three-site cluster. At half filling, the low-energy manifold on each rung is spanned by two nearly degenerate states with one and two bosons, respectively. These two states can be mapped onto an effective pseudospin-$\tau = 1/2$ system. Here, the low-energy manifold on each rung is spanned by the states $\ket{\uparrow_r} = \ket{101_r}$ and $\ket{\downarrow_r} = \ket{010_r}$, which we identify as the basis of pseudospin-operators ($\tau^{+}_r, \tau^{-}_r, \tau^z_r$). In this notation, $\ket{n_{r,1}, n_{r,2}, n_{r,3}}$ represents the Fock basis with occupation numbers on the three sites of rung $r$. Projecting the inter-rung hopping onto this subspace, the first-order contribution yields an effective XY-type coupling,
\begin{align}
    H_{\mathrm{eff}}^{(1)}
    =
    -J_{xy}^{(1)}(\phi)\sum_r
    \left(
    \tau_r^+ \tau_{r+1}^- + \tau_r^- \tau_{r+1}^+
    \right),
\end{align}
with
\begin{align}
    J^{(1)}_{xy}(\phi) = J_{\parallel}\left(\cos\phi + \frac14\right).
\end{align}
This term naturally accounts for the superfluid-like phases in the phase diagram. In particular, the sign change of $J^{(1)}_{xy}(\phi)$ (Fig.~\ref{fig:perturbation}(a)) provides a simple interpretation of the transition between the Meissner and reverse-Meissner regimes: for $J^{(1)}_{xy}(\phi) > 0$, the effective XY coupling favors the conventional Meissner-like current pattern, whereas for $J^{(1)}_{xy}(\phi) < 0$, it favors the opposite chirality, corresponding to the reverse-Meissner phase. Thus, the first-order effective Hamiltonian already explains why the direction of the chiral current changes as a function of $\phi$.

More importantly, at second order in $J_{\parallel}/J_{\perp}$, a diagonal interaction is generated in the effective pseudospin model:
\begin{align}
    H_{\mathrm{eff}}^{(2)} = -J^{(2)}_z(\phi)\sum_r \tau_r^z \tau_{r+1}^z + \, \text{const},
\end{align}
with
\begin{align}
    J^{(2)}_z(\phi) = \frac{J_{\parallel}^2}{J_\perp^{}} \left( \frac{3}{2} \cos(\phi)^2 + \frac{3}{4} \cos\phi - \frac{119}{192} \right).
\end{align}
Thus, although the original Hamiltonian does not contain an explicit density-density interaction along the ladder direction, the nearest-neighbor interaction emerges through 2nd-order perturbative hopping processes. Since $\tau^z$ corresponds to the imbalance between the two low-energy rung occupations, antiferromagnetic ordering in the pseudospin language directly corresponds to a CDW pattern such as ($\ket{101_1}\ket{010_2}\ket{101_3} ...\ket{101_L}$). The perturbative expression for $J_z^{(2)}(\phi)$ indicates that it becomes positive in a finite window of $\phi$, which is broadly consistent with the region where the CDW phase is observed numerically. This suggests that the CDW is not simply an accidental feature of the numerics, but can be understood as an emergent interaction-driven instability within the low-energy manifold.

\begin{figure}[t]
\centering
\includegraphics[width=1.0\linewidth]{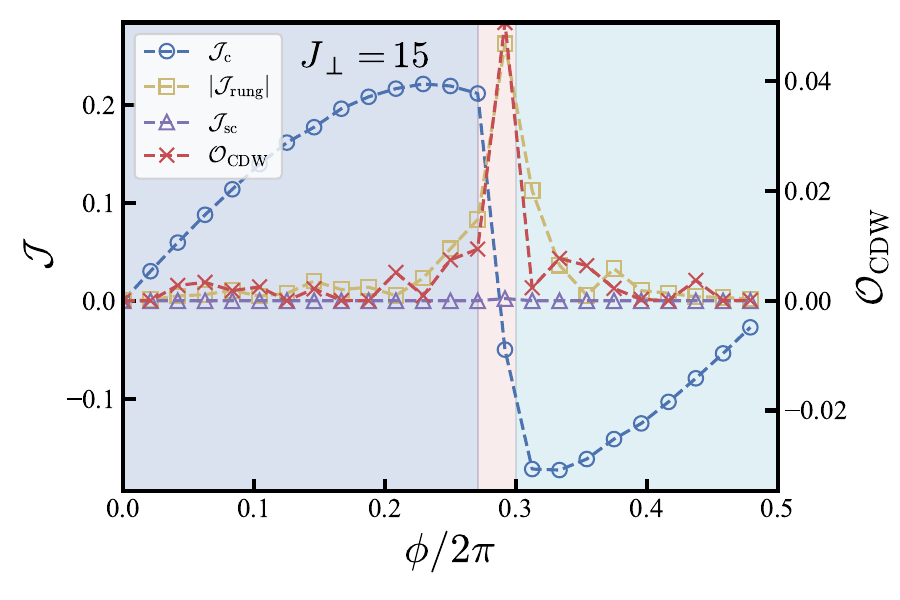}
\caption{
Flux dependence of the effective couplings the chiral current $\mathcal{J}_c$, the rung current amplitude $|\mathcal{J}_{\mathrm{rung}}|$, the staggerd current $\mathcal{J}_{\text{sc}}$, and the CDW order parameter $\mathcal{O}_{\mathrm{CDW}}$ (right axis) at $J_{\perp}=15$. The blue region is the M-SF phase, where ${\mathcal J}_c$ increases monotonically with $\phi$. The red region is the CDW state where ${\mathcal O}_{\mathrm{CDW}}$ is sharply enhanced. The light blue region is the RM-SF phase, where ${\mathcal J}_c$ has the opposite sign to that in the M-SF phase. Note that the red region cannot be simply identified as a CDW phase, since it also shows vortex-like features (see the rung current $|\mathcal{J}_{\text{rung}}|$).
}
 \label{fig:phase_diagram_Jr15}
\end{figure}

Interestingly, the perturbative expression also shows that $J_z^{(2)}(\phi)$ diverges as $\phi$ approaches the point where $J_{xy}^{(1)}(\phi)$ vanishes (Fig.~\ref{fig:perturbation}(a),(b)). However, this divergence should not be interpreted literally. Near this point, the second-order contribution becomes comparable to or even exceeds the first-order term, signaling the breakdown of the perturbation theory. Therefore, the perturbative treatment cannot be trusted quantitatively in the immediate vicinity of this singular point. However, the effective theory for our systems provides useful qualitative insight. Although the original microscopic Hamiltonian does not contain an explicit density-density interaction, second-order virtual processes naturally generate an effective CDW-inducing interaction $J_z$. This may be relevant to understanding the emergence of the CDW order. Indeed, around the singular point $\phi \simeq 0.3$, our numerical results at $J_{\perp}=15$ show that the CDW order appears (Fig.\ref{fig:phase_diagram_Jr15}). This result is consistent with the tendency suggested by the perturbation theory.


In the intermediate regime $J_{\perp}\simeq J_{\parallel}$, the mechanism stabilizing the CDW phase is not fully understood. A general effect of the artificial gauge field is to suppress uniform phase coherence, thereby promoting vortex formation. In conventional ladder systems, this typically leads to vortex-dominated superfluid phases characterized by non-uniform, twisted phase coherence. In the present system, however, once the uniform phase coherence is weakened, the system does not simply enter a vortex phase. Instead, vortex ordering competes with a density-ordered state, leading to the emergence of the CDW phase. The key of this nontrivial phenomenon is that the CDW order becomes strong enough to compete with vortex states on comparable energy scales over a wide parameter region. 
This indicates that the CDW phase is not a perturbative instability of a simple reference state, but an emergent phenomenon in the intermediate regime. Such competition is absent in two-leg ladders, highlighting the essential role of the three-leg geometry. The microscopic origin of this strong enhancement of the density ordering remains unclear and cannot be captured by a simple perturbative argument. In particular, the detached island-like CDW region is connected to neither the strong rung- nor leg-coupling limits. This suggests that the CDW phase is not a trivial consequence of effective interactions, but rather an emergent phenomenon arising from the interplay of frustration and quantum fluctuations.

Our results therefore establish the CDW phase as a robust numerical finding in a regime where competing orders coexist, and call for further theoretical understanding beyond the present analysis.

\section{Summary}
In this work, we studied a three-leg hard-core Bose-Hubbard ladder at half filling under an artificial gauge field. By systematically analyzing correlation functions and current patterns, we reveal the rich phase diagram containing multiple superfluid states. We also found that CDW order emerges over a broad parameter region where vortex states are naturally expected and indeed realized in the neighborhood, despite the presence of only on-site interactions. Although part of this behavior can be qualitatively understood from a perturbative analysis in the strong rung-coupling limit, the full origin of the CDW phases remains nontrivial. In particular, neither the persistence of CDW order in the regime where rung and leg coupling compete nor the appearance of an isolated island-like CDW region can be straightforwardly accounted for within the simple strong-coupling picture. Our results therefore reveal that artificial gauge fields and strong correlations can cooperate to produce unexpectedly robust and unconventional density-ordered phases in bosonic ladder systems which emerge in a flux regime where vortex states are conventionally expected.

Our results demonstrate that few-leg bosonic ladders provide a minimal and versatile platform for exploring unconventional gauge flux responses, such as vortex physics and reentrant ordering phenomena. These findings could open new perspectives on how artificial gauge field cooperate to generate novel many-body ground states beyond conventional superfluid paradigms.

\begin{acknowledgments}
We thank Shun Uchino for fruitful discussions. The DMRG calculations were performed with the use of ITensor.jl \cite{ITensor}.  Numerical calculations have been done using the facilities of the Supercomputer Center, the Institute for Solid State Physics, the University of Tokyo. This study is supported by JST SPRING Grant No JPMJSP2132 and JSPS KAKENHI Grant No. 22K03513.

\end{acknowledgments}

\appendix
\section{Constraint on generation of mass gap from the Oshikawa-Yamanaka-Affleck condition}
\label{app:OYA}
We discuss the absence of the Mott-insulating state characterized by the mass gap
\begin{equation}
\Delta E_{\mathrm{mass}} = E(N+1) + E(N-1) - 2E(N),
\end{equation}
where $E(N)$ denotes the ground-state energy in the sector with $N$ particles. In ladder systems, the uniquely gapped state is constrained by the Oshikawa--Yamanaka--Affleck (OYA) condition \cite{PRL.78.1984}. For a spin-$S$ system, the OYA condition is given by
\begin{equation}
l_{\text{unit}}\, \mathcal{N} (S - \langle S_z \rangle) \in \mathbb{Z},
\end{equation}
where $l_{\text{unit}}$ is the length of the (enlarged) unit cell and $\mathcal{N}$ is the number of legs. When this condition is satisfied for $l_{\text{unit}}=1$, the system can support a magnetization plateau with translation symmetry, which in the present bosonic context corresponds to a uniquely gapped Mott-insulating phase.

In our model, the mapping yields $N=3$, $S=1/2$, and $\langle S_z \rangle = \langle n \rangle - 1/2$, such that the magnetization plateau can be reinterpreted as the opening of a mass gap in the particle-number sector. At one-third filling $\langle n\rangle=1/3$, the OYA condition can be satisfied for arbitrary gauge flux, allowing the ground state to be a uniquely gapped Mott-insulating phase \cite{Kolley_2015}. In contrast, at half filling $\langle n\rangle=1/2$, this condition is not satisfied for any choice of parameters and therefore the ground state must be non-uniquely gapped or gapless.

\section{Discussion about stability of CV-SF phase against finite-size effects}
\label{app:finite_size}

\begin{figure*}[t]
  \centering
  \includegraphics[width=1.0\linewidth]{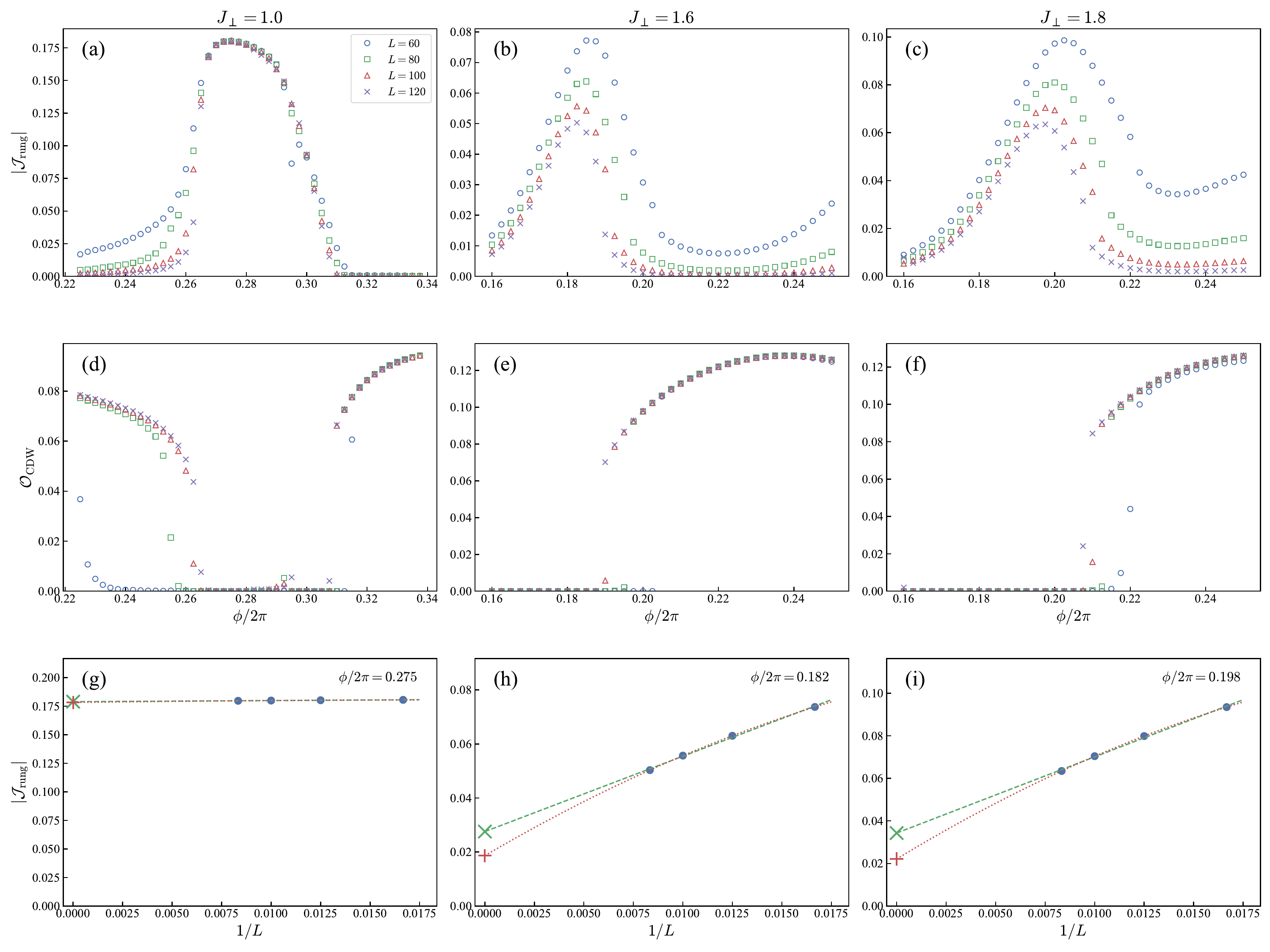}
  \caption{
Finite-size dependence of the CV-SF regions. (a)--(c) The rung-current amplitude $|\mathcal{J}_{\rm rung}|$ for $J_\perp =1.0$, $1.6$, and $1.8$, respectively. (d)--(f) The CDW order parameter $\mathcal{O}_{\rm CDW}$. (g)--(i) Finite-size extrapolation of $|\mathcal{J}_{\rm rung}|$ as a function of $1/L$ at representative fluxes, $\phi/2\pi=0.275$, $0.182$, and $0.198$, respectively. The green dashed line ('x' markers) shows the linear fit, and the red dashed line ('+' markers) shows the second-order polynomial fit. For $J_\perp =1.0$, the finite rung current shows only weak system-size dependence, supporting the stability of the CV-SF region in the thermodynamic limit. In contrast, for $J_\perp =1.6$ and $1.8$, the magnitude of $|\mathcal{J}_{\rm rung}|$ is strongly suppressed with increasing system size, indicating that these narrow CV-SF regions are sensitive to finite-size effects.
}
\label{fig:finite_size}
\end{figure*}
\begin{figure*}[t]
  \centering
  \includegraphics[width=1.0\textwidth]{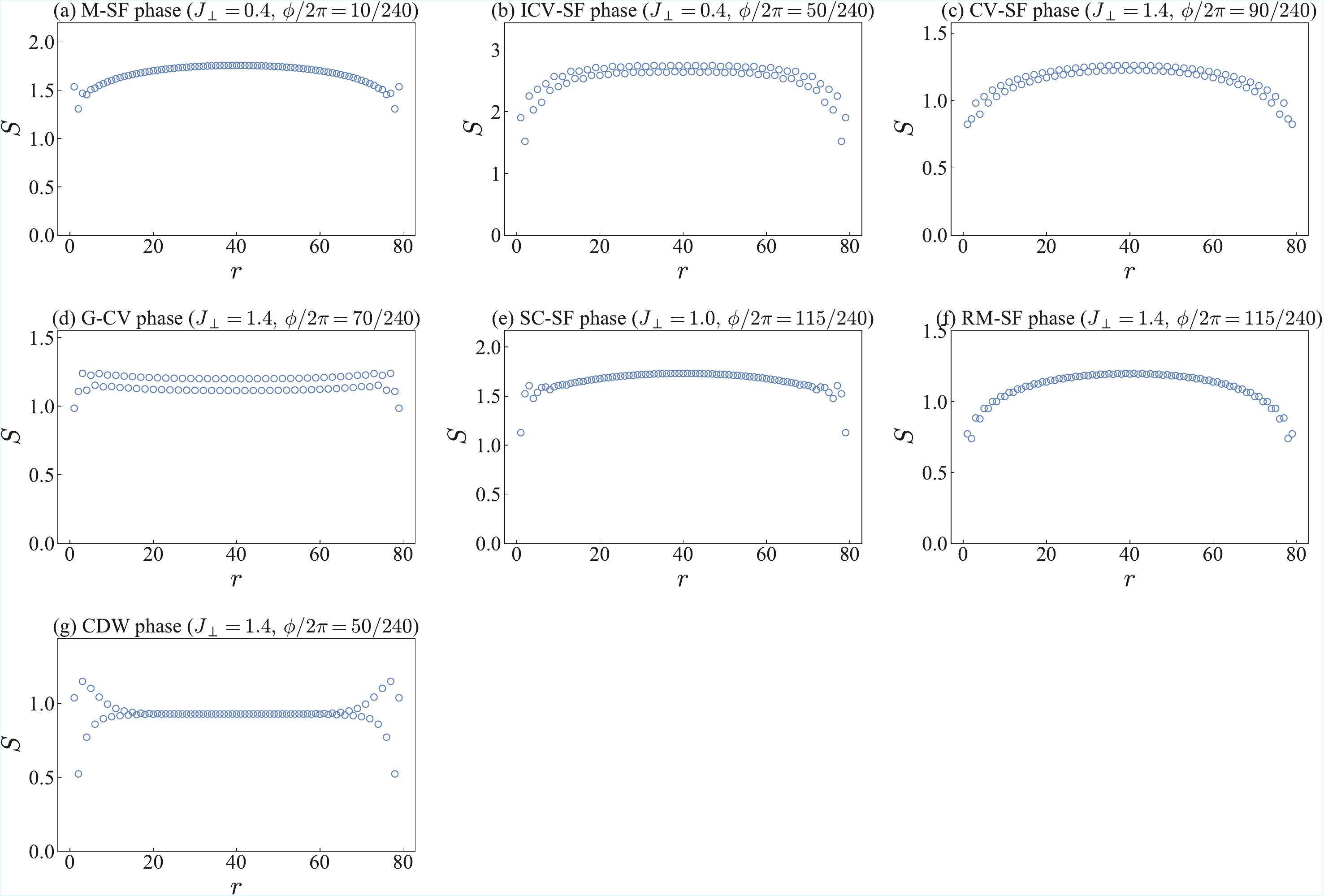}
  \caption{Entanglement entrooy for representative parameters in different quantum phases. Panels (a)-(g) correspond to the M-SF, ICV-SF, CV-SF, G-CV, SC-SF, RM-SF and CDW, respectively. Distinct behaviors of entanglement entropy highlight qualitative differences between gapless (superfluid) phases and gapped phases.}
  \label{fig:central_charge}
\end{figure*}

In the main text, the phase diagram in Fig.~\ref{fig:phase_diagram} is determined mainly from DMRG calculations for the system size $L=80$. Since the phase diagram contains several narrow regions of CV-SF phase, it is important to discuss whether these phases are robust against finite-size effects. In this Appendix, we present additional DMRG results for different system sizes along representative cuts of the phase diagram. We perform calculations for $L=60,80,100$ and 120 and compare the flux dependence of the order parameters used in the main text.

Figure~\ref{fig:finite_size}(a)(d) shows the finite-size dependence of several physical quantities for $J_\perp=1.0$. These cuts are chosen to examine whether the narrow CV-SF regions survive against finite-size effects. In both cases, the CDW order parameter $\mathcal{O}_{\rm CDW}$ (Eq.~\eqref{eq:CDW}) and the rung-current amplitude $|\mathcal{J}_{\rm rung}|$ (Eq.~\eqref{eq:Jr}) show only weak system-size dependence for $L=60,80,100$ and $120$. For $J_\perp=1.0$, $\mathcal{O}_{\rm CDW}$ is suppressed in an intermediate flux region where $|\mathcal{J}_{\rm rung}|$ becomes finite, and it reappears at larger flux, corresponding to the reentrant phase transition discussed in the main text (Sec.~\ref{sec:CDW}).  These results indicate that the intermediate CV-SF regions and reentrant CDW phase transition are not finite-size artifacts of the DMRG calculations.

In contrast to the $J_\perp \simeq 1.0$ case, we find that the other narrow vortex region in the phase diagram are sensitive to finite-size effects. In particular, for $J_\perp=1.6$ and $1.8$ in the range $1.5 \lesssim \phi \lesssim 2.0$, a small region with finite $|\mathcal{J}_{\rm rung}|$ and suppressed $\mathcal{O}_{\rm CDW}$ appears in finite-size systems. However, calculations for $L=60,80,100,$ and $120$ show that the magnitude of $|\mathcal{J}_{\rm rung}|$ in this region decreases as the system size is increased (Fig.~\ref{fig:finite_size}(b)(c)(e)(f)). To estimate the thermodynamic-limit value of the rung-current amplitude, we fit the finite-size data to linear and second-order polynomial functions in $1/L$. As shown Fig.~\ref{fig:finite_size}(g)-(i), in contrast to the result of extrapolation for $J_\perp = 1.0$, the data do not provide clear evidence that this vortex region survives in the thermodynamic limit for $J_\perp = 1.6$ and 1.8. Therefore, unlike the robust CV-SF phase near $J_\perp =1.0$, we regard this region as a unstable CV-SF region in thermodynamics limit.

We also examine the nature of the phase transition between the CDW and CV-SF regions in more detail. As shown in Fig.~\ref{fig:finite_size}(d), the CDW order parameter $\mathcal{O}_{\rm CDW}$ change rather abruptly near the phase boundary. This behavior is seen for different system sizes and is suggestive of a first-order-transition-like transition. Therefore, the phase transitions between the CDW and CV-SF phases do not appear to be continuous transitions, but rather reflect a direct competition between two distinct ordered states.

\section{Point group symmetry}
\label{app:pg}

There are $C_2$ and mirror symmetries in the Hamiltonian~\eqref{eq:Hamiltonian} under the gauge filed~\eqref{eq:vector_potential}~\cite{Tada_2026}. Since the system considered in the DMRG calculation has the length $L=80 (r=1,2,\cdots,L)$ with the open boundary condition, there exists a $C_2$ rotation symmetry about the bond connecting the sites $(r,\ell)=(L/2,2), (L/2+1,2)$ which acts as $C_2:(r,1)\to(L-r+1,3), (r,2)\to(L-r+1,2)$. The mirror symmetry $M$ flips the flux $\phi\to-\phi$ and it is not a symmetry of the Hamiltonian by itself. (Note that there are two mirrors, namely $M_x:(r,\ell)\to (L-r+1,\ell)$ and $M_y:(r,1)\to(r,3), (r,2)\to(r,2)$. ) Thus, we combine $M$ with the particle-hole conjugation $\mathcal{C}:b_{r,\ell}\to b_{r,\ell}^{\dagger}$ and define $\tilde{M}=\mathcal{C}M$. The combined mirror operators commute with the Hamiltonian in the hard-core limit $U\to\infty$. Furthermore, they also commute with the $C_2$ rotation, $[C_2,\tilde{M}_{x,y}]=[\tilde{M}_x,\tilde{M}_y]=0$.

Under the $C_2$ rotation, the current operators are transformed as $C_2:j_{r,1}^{\parallel}\to -j_{r',3}^{\parallel}, j_{r,2}^{\parallel}\to -j_{r',2}^{\parallel}$ and $C_2:j_{r,1}^{\perp}\to -j_{r',2}^{\perp}$, where $r'=L-r+1$. Similarly, $\tilde{M}_x:j_{r,\ell}^{\parallel}\to +j_{r',\ell}^{\parallel}$ and $\tilde{M}_x:j_{r,\ell}^{\perp}\to -j_{r',\ell}^{\perp}$ with $r'=L-r+1$. Also, $\tilde{M}_y:j_{r,1}^{\parallel}\to -j_{r,3}^{\parallel}, j_{r,2}^{\parallel}\to -j_{r,2}^{\parallel}$ and $\tilde{M}_y:j_{r,1}^{\perp}\to +j_{r,2}^{\perp}$.

The ground state can be a simultaneous eigenstate of all the commuting symmetries. Therefore, the current patterns in such a ground state satisfy symmetry constraints corresponding to the above transformation properties of the current operators. However, in the DMRG calculations, we do not use the quantum numbers associated with the point group symmetry and hence the symmetry properties of the current patterns may be broken numerically. Indeed, the ground state obtained by DMRG for the SC-SF phase explicitly breaks the $\tilde{M}_y$ symmetry (Fig.~\ref{fig:schematic}).

\section{Discussion about entanglement entropy and central charge}
\label{app:enta}

We briefly discuss the entanglement entropy.  We consider the entanglement entropy of the system under a bipartition into the right subsystem A and the left subsystem B:
\begin{align}
S(r) = - \mathrm{Tr_{B}} \left[ \rho_\mathrm{A} \mathrm{log}\rho_\mathrm{A} \right],
\label{eq:entanglement}
\end{align}
where $r$ is length of the subsystem A and $\rho_A$ is reduced density matrix of subsystem A. When the system is gapless, the entanglement entropy follows Calabrese-Cardy formula \cite{Calabrese_2004},
\begin{align}
S(r)  = \frac{c}{6} \mathrm{log} \left[ \frac{2 L}{\pi} \mathrm{sin} \left( \frac{\pi r}{L} \right) \right] + \mathrm{const}, 
\label{eq:centralcharge}
\end{align}
where $c$ is the central charge, which roughly corresponds to the number of gapless excitation modes under U(1) symmetry. In contrast, for a gapped system, there are no gapless modes and the system effectively behaves as if $c=0$, leading to a nearly flat dependence of $S$ on $r$. 
These distinct behaviors allow us to determine whether the system is gapped or gapless from the scaling of the entanglement entropy. 

Figure~\ref{fig:central_charge} shows the entanglement entropy evaluated for representative parameters in the different quantum phases identified in the phase diagram. In the superfluid phases, including the M-SF phase, ICV-SF phase, SC-SF phase, CV-SF phase and RM-SF phases, the entanglement entropy exhibits a characteristic smooth arc-like profile, consistent with the logarithmic scaling expected for gapless one-dimensional critical states. In contrast, the CDW phase and G-CV phase show a qualitatively different behavior. These phases exhibit a nearly flat entanglement entropy profile, consistent with a gapped phase. 

%

\end{document}